# Observational evidence of anisotropic changes apparent resistivity before strong earthquakes

*Jianguo Zhang[1,2,*] Wei Du[2] Mingxin Yue[2] Chenghui Liu[2] Xiaolong Liang[2] and Jun Yang[2]*
1 *Handan Seismostation of Seismological Bureau of Hebei Province, Handan 056001, China*
2 *School of Earth and Space Sciences, University of Science and Technology of China, Hefei 230026, China*

**Summary**

Using a method based on normalized monthly variation rate, we studied resistivity data of seven observation stations before the events in the epicenter areas of two strong earthquakes. The relationship between variation of anisotropic apparent resistivity and the azimuth of the maximum principal stress is analyzed. The study shows that significant apparent resistivity variation occurs in the direction that is perpendicular to the azimuth of the maximum principal stress while only small fluctuation are recorded in the direction of the maximum principal stress. We surmise that the variation of anisotropic resistivity occurs in the late stage of the development of a strong earthquake, which can be observed in the epicenter area. If the density of the observation stations is increased and the direction of the observed resistivity is right, the epicenter of an earthquake location may be estimated by the observed resistivity anomaly.

**Introduction**

Short-term earthquake prediction still remains one of the most challenging and debated questions in the scientific community. Nevertheless, earthquake-related electromagnetic phenomena have recently been considered as a one of the most promising candidates to achieve short-term earthquake prediction. The electromagnetic change is one of the most promising phenomena in this regard suggesting that short-term prediction is expected to be feasible.

In this paper, using the normalized monthly rate method, we study the variation of resistivity anisotropy and focal mechanism solutions the maximum principal stress orientation (p-axis direction) correspondence on strong earthquakes. So, it can help us set up ground resistivity stations and select the buried electrode orientation.

**Selection of observation data and earthquakes**

The instrument used in this study is ZD8 digital earth-resistivity observation instrument, which is produced by Lanzhou Institute of Seismology, China Seismological Bureau, the instrument sampling rate of the hour, the electrode resistivity surveying instrument layout mode most of the NS and, EW two measurement channels direction. We select seven resistivity stations near the Wenchuan and Yunshu earthquake, and used all observations since they have been established.

Table 1: Information parameters of Wenchuan and Yushu earthquakes

| Epicenter | Time | Magnitude (Ms) | Longitude (°) | Latitude (°) | Depth (KM) |
|---|---|---|---|---|---|
| WenChuan | 2008.5.12 | 8.0 | 103.4 | 31.0 | 10 |
| YuShu | 2010.4.14 | 7.1 | 96.7 | 33.1 | 14 |

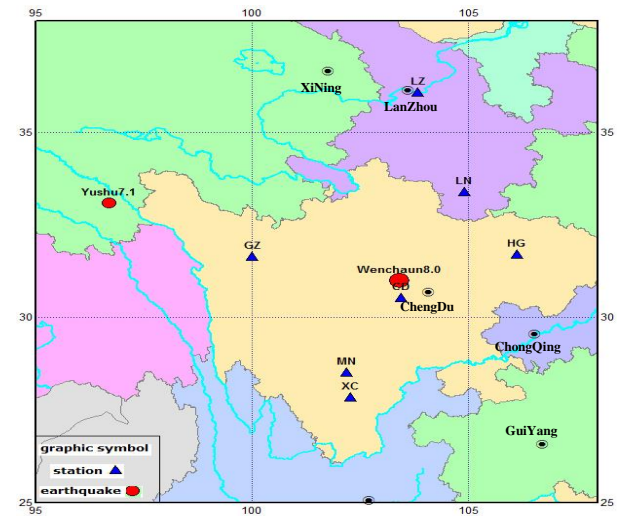

Figure 1: The earthquakes epicenter and resistivity stations location

**Normalized monthly variation ratio method**

When extracting anomaly by the method of normalized monthly variation ratio, first of all, make data download, exclude annual and non-annual periodic variations, make periodic processing to the data with periodic and long trends. We then calculate the slope of monthly average (five-day average, daily average) curve relative to the

## Methodologies and Technologies

timeline with a certain step and make normalized or relative change processing. The calculation process is as follows:

The slope of observed quantities curve relative to the timeline is given by:

$$k_i = \frac{\sum_{j=1}^{n}T_j \sum_{j=1}^{n}y_j - n\sum_{j=1}^{n}T_j y_j}{(\sum_{j=1}^{n}T_j)^2 - n\sum_{j=1}^{n}T_j^2} \quad (1)$$

Autocorrelation coefficients:

$$R_i = \frac{\sum_{j=1}^{n}T_j y_j - \frac{1}{n}(\sum_{j=1}^{n}T_j \sum_{j=1}^{n}y_j)}{\sqrt{[\sum_{j=1}^{n}T_j^2 - \frac{1}{n}(\sum_{j=1}^{n}T_j)^2]\cdot[\sum_{j=1}^{n}y_j^2 - \frac{1}{n}(\sum_{j=1}^{n}y_j)^2]}} \quad (2)$$

The sequence of normalized monthly variation ratio method:

$$S_i = R_i \times K_i / \sigma_{n-1} \qquad (i = n, n+1, ..., N) \quad (3)$$

where n denotes slide step, N denotes data length, $\sigma_{n-1}$ denotes the mean square error (MSE) of N-n $R_i \times K_i$ time series, $\{y\}$ denotes equally spaced time series of precursor data, $\{T\}$ denotes corresponding equally spaced time series. $S_i$ denotes month rate when using monthly average curve slide; $S_i$ denotes five-day rate of change when using five-day average curve slide. Extracting short-term anomaly, it is timesaving and convenient to calculate month rate with monthly average curvy, and anomaly change is prominent, more significantly.

**Analysis of calculating result**

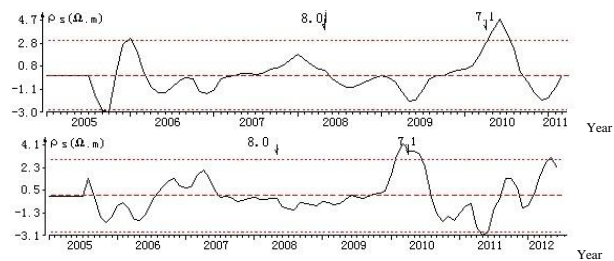

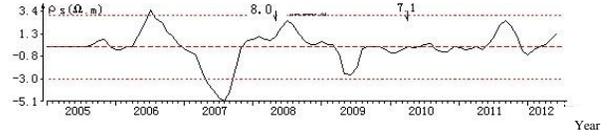

Figure 2: Apparent resistivity monthly rate change curve of Lanzhou station

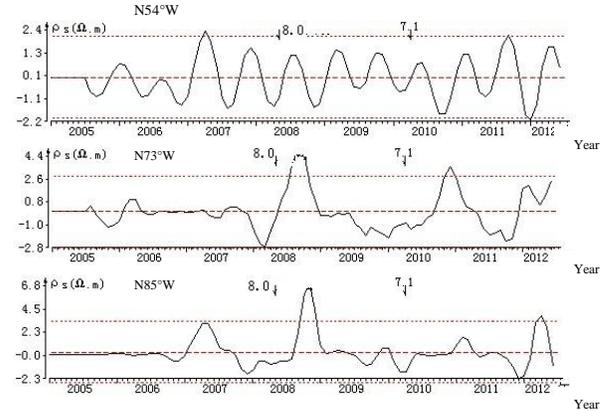

Figure 3: Apparent resistivity monthly rate change curve of Longnan station

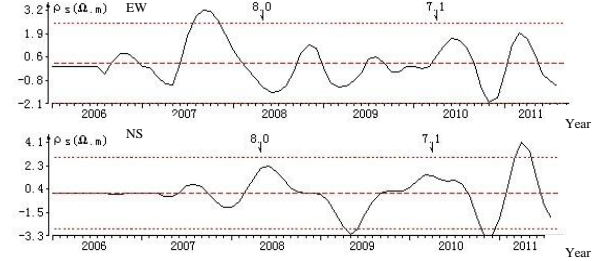

Figure 4: Apparent resistivity monthly rate change curve of Chengdu station

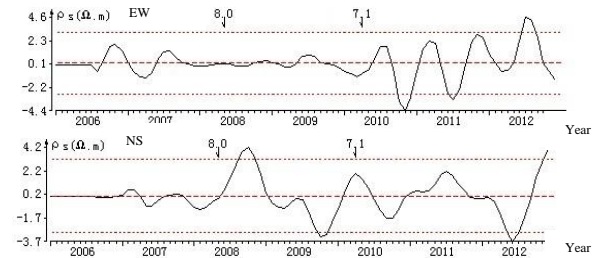

Figure 5: Apparent resistivity monthly rate change curve of Xichang station

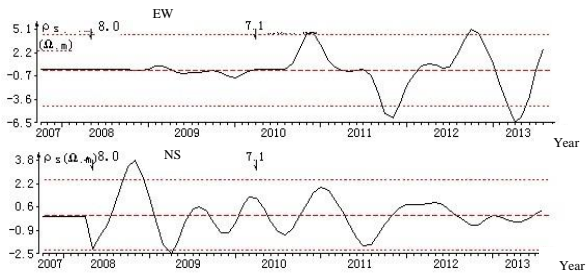

Figure 6: Apparent resistivity monthly rate change curve of Ganzi station

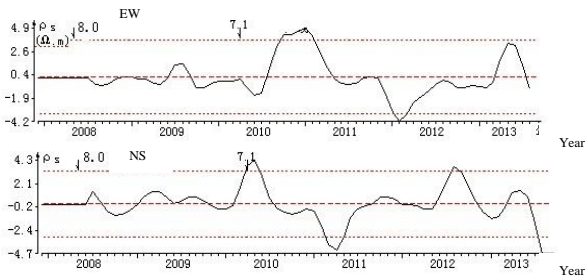

Figure 7 Apparent resistivity monthly rate change curve of Hongge station

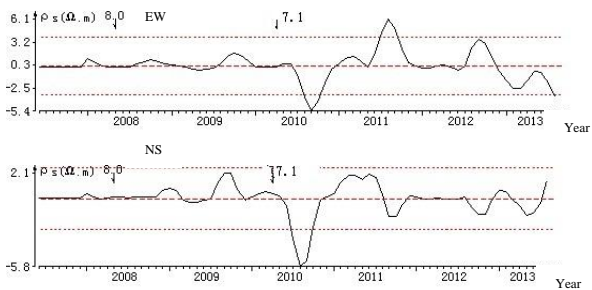

Figure 8: Apparent resistivity monthly rate change curve of Mianning station

(1) The fall and rise of impending change is more obvious in the time scale of one year most of the stations before the earthquakes;
(2) Defined |S|≥2.4 as an exception criterion, when most of the seismostations (during Wenchuan earthquake) occur impending abnormal changes, the abnormal finding show magnitude different in different observed direction, which reflect electrical sensitivity, ambient change and other differences exist in the underground medium of the different stations.

**Conclusions**

The results show that in the medium short and short imminent terms(in one year before the earthquake) of earthquake preparation, the variation rate perpendicular (or near perpendicular) to the direction of maximum principal compression stress is greater than that parallel (or near parallel) to the direction.

The anisotropic changes in apparent resistivity from most stations (in Wenchuan earthquake) are related to the maximum compressive principal stress obtained from focal mechanism solutions in the late incubation period of earthquake. Namely, for the same station, it changes sharply when the monitoring channel is perpendicular (or near perpendicular) to the direction of maximum principal compression stress, while it changes slightly (or not up to anomaly evaluation criteria) when the channel is parallel (or near parallel) to the direction, which accords with the anisotropic variations in apparent resistivity of most loading process of rock experiments.

Based on the analysis of the mechanism of anisotropic changes in apparent resistivity, in the late incubation period of strong earthquake, squeezing action has a significant effect near the direction of maximum principal compression stress near the focal region, the vertical micro-cracks develop fast in number and size, then strike predominantly along the maximum compressional stress orientation, while conductive fluids enter fleetly or re-distribute, then result in sharp and fast variations in true resistivity along the maximum loading direction.

**Acknowledgements**

This research is supported by the National Natural Science Foundation of China (No.41274079) and the Key Research Project of Seismological Bureau of Hebei Province ([2010]20). I appreciate the two anonymous reviewers for their constructive comments on the original manuscript paper and also thank my research teams.